\documentclass[11pt,a4paper]{article}

\usepackage[utf8]{inputenc}
\usepackage[T1]{fontenc}
\usepackage{lmodern}
\usepackage{microtype}
\usepackage[margin=1in]{geometry}
\usepackage{amsmath,amssymb,amsthm}
\usepackage{listings}
\usepackage{xcolor}
\usepackage{booktabs}
\usepackage{enumitem}
\usepackage{graphicx}
\usepackage{caption}
\usepackage{subcaption}
\usepackage{algorithm}
\usepackage{algpseudocode}
\usepackage{tikz}
\usetikzlibrary{arrows.meta,positioning,shapes.geometric,fit}
\usepackage[hyphens]{url}
\usepackage[most]{tcolorbox}
\usepackage{float}
\usepackage[section]{placeins}
\usepackage[colorlinks=true,linkcolor=blue!70!black,citecolor=green!50!black,urlcolor=blue!70!black,bookmarks=true]{hyperref}

\tcbset{
  turnbox/.style={
    enhanced,
    arc=2mm,
    boxrule=0.7pt,
    left=8pt,right=8pt,top=7pt,bottom=7pt,
    fonttitle=\bfseries\small,
  }
}

\definecolor{turnbg}{RGB}{248,248,248}
\definecolor{turnkw}{RGB}{0,100,180}
\definecolor{turnstr}{RGB}{180,60,30}
\definecolor{turncmt}{RGB}{100,140,100}
\definecolor{turntype}{RGB}{140,50,140}

\lstdefinelanguage{Turn}{
  keywords={let,struct,turn,infer,spawn,spawn_link,spawn_each,send,receive,self,return,if,else,while,try,catch,throw,use,grant,suspend,remember,recall,call,context,confidence,map,filter,null,true,false,and,or},
  keywordstyle=\color{turnkw}\bfseries,
  ndkeywords={Str,Num,Bool,List,Map,Pid,Vec,Identity},
  ndkeywordstyle=\color{turntype}\bfseries,
  sensitive=true,
  comment=[l]{//},
  morecomment=[s]{/*}{*/},
  string=[b]",
  stringstyle=\color{turnstr},
  commentstyle=\color{turncmt}\itshape,
  basicstyle=\ttfamily\footnotesize,
  breaklines=true,
  showstringspaces=false,
  tabsize=4,
  frame=tb,
  framerule=0.4pt,
  framesep=5pt,
  backgroundcolor=\color{turnbg},
  xleftmargin=4pt,
  xrightmargin=4pt,
  numbers=none,
  aboveskip=6pt,
  belowskip=4pt,
}

\newtheorem{definition}{Definition}

\newtheorem{property}{Property}

\usepackage{fancyhdr}
\fancypagestyle{plain}{
  \fancyhf{}
  \fancyhead[L]{\small\textit{Turn: A Language for Agentic Computation}}
  \fancyhead[R]{\small\textit{Preprint}}
  \fancyfoot[C]{\thepage}

}

\begin{document}

\twocolumn[{%
\begin{center}
  \vspace{0.4em}
  {\LARGE\bfseries Turn: A Language for Agentic Computation\par}
  \vspace{0.35cm}
  {\large Muyukani Kizito\,$^{1,2}$\par}
  \vspace{0.15cm}
  {\normalsize $^1$Turn Lang \quad $^2$Prescott Data\par}
  \vspace{0.3cm}
\end{center}
\begin{center}
  \textbf{\large Abstract}
\end{center}
\vspace{0.2em}
We present \textbf{Turn}, a compiled, actor-based programming language---statically typed for schema inference, dynamically typed at the value level---for agentic software: programs that reason and act autonomously by delegating inference to large language models (LLMs). Existing approaches augment general-purpose languages with frameworks, encoding critical invariants (bounded context, typed inference output, credential isolation, durable state) as application-level conventions rather than language guarantees.

Turn introduces five language-level constructs that address this gap. \emph{Cognitive Type Safety} makes LLM inference a typed primitive: the compiler generates a JSON Schema from a struct definition and the VM validates model output before binding. The \emph{confidence operator} enables deterministic control flow gated on model certainty. Turn's \emph{actor-based process model}, derived from Erlang, gives each agent an isolated context window, persistent memory, and mailbox. A \emph{capability-based identity system} returns opaque, unforgeable handles from the VM host, ensuring raw credentials never enter agent memory. Finally, \emph{compile-time schema absorption} (\texttt{use schema::<protocol>}) synthesizes typed API bindings from external specifications at compile time; the \texttt{openapi} adapter is shipped with \texttt{graphql}, \texttt{fhir}, and \texttt{mcp} in active development.

We describe the language design, type rules, schema semantics, and a Rust-based bytecode VM, and evaluate Turn against representative agentic workloads. Turn is open source at \url{https://github.com/ekizito96/Turn}.
\vspace{1em}
\vspace{0.5em}


\vspace{0.9em}
}]

\section{Introduction}
\label{sec:introduction}

The emergence of large language models (LLMs) as general-purpose reasoning engines has produced a new class of software: \emph{agentic programs} that observe their environment, reason over unstructured data, invoke external tools, and act on their conclusions. Unlike traditional software, where all behavior is deterministic, agentic programs delegate critical decision-making to stochastic models whose outputs are structurally unpredictable.

Building these systems with existing languages and frameworks reveals a fundamental impedance mismatch. The core abstractions of agentic computation---bounded context windows, typed inference results, durable execution state, process-isolated memory, and credential security---have no direct representation in languages like Python, TypeScript, or Rust. Developers encode them as library conventions, leading to five recurring failure modes:

\begin{enumerate}[leftmargin=2em]
    \item \textbf{Unbounded context.} Context is managed as a plain list of messages. When it exceeds the model's token limit, it overflows silently or truncates from the front, causing the model to lose critical information.
    \item \textbf{Untyped inference output.} The model returns a string. The program parses it as JSON with no guarantee that the schema matches. Schema violations surface as runtime key errors with no connection to the inference that produced them.
    \item \textbf{Fragmented state.} Agent state is scattered across variables, databases, API responses, and framework internals. There is no single, serialisable representation of what the agent knows or has done.
    \item \textbf{No durable execution.} Long-running agents die with their process. There is no language-level mechanism for checkpointing and resuming execution across restarts.
    \item \textbf{Credential leakage.} API keys and OAuth tokens are stored as strings accessible to the LLM's tool-calling interface, creating an exfiltration vector in any system where the model can invoke I/O.
\end{enumerate}

These failure modes are not theoretical. They arose in practice during the development of \textsc{JarvisCore} and other production agentic systems at Prescott Data, where empirical analysis \cite{kizito2026physics,kizito2026subagent} confirmed that reliability at scale requires language-level enforcement of the invariants that frameworks leave as convention. Turn is the result: a language in which each of these constraints is a first-class primitive rather than a library pattern.

\paragraph{Paper structure.} \S\ref{sec:related} surveys related work; readers already familiar with the actor model, capability security, and LLM output structuring may proceed directly to \S\ref{sec:design}. The contributions in \S\ref{sec:cognitive-type-safety}--\S\ref{sec:schema-absorption} are self-contained and can be read in any order after \S\ref{sec:design}.

\subsection{Contributions}

This paper makes the following contributions:

\begin{enumerate}[leftmargin=*]
    \item \textbf{Cognitive Type Safety} (\S\ref{sec:cognitive-type-safety}): We introduce \texttt{infer Struct \{ prompt \}}, a typed inference primitive where the compiler generates a JSON Schema from the struct definition at compile time and the VM validates the model's response before binding. We prove that if an \texttt{infer} expression completes without error, the bound value structurally conforms to the declared type.

    \item \textbf{Probabilistic Control Flow} (\S\ref{sec:probabilistic-routing}): We introduce the \texttt{confidence} operator, a first-class language construct that extracts the model's reported certainty from an inference result and enables deterministic branching on stochastic output.

    \item \textbf{The Agentic Process Model} (\S\ref{sec:actor-model}): We adapt Erlang's actor model to the agentic domain. Rather than a flat list of messages, each agent process has a tripartite \emph{structured context architecture} to enforce high-recall invariants, O(1) persistent agent memory, an asynchronous mailbox, and \emph{durable execution} via exact \texttt{suspend}/\texttt{resume} checkpoints.

    \item \textbf{Capability-Based Identity} (\S\ref{sec:identity}): We introduce \texttt{grant identity}, a primitive that returns an opaque, unforgeable \texttt{Identity} capability handle from the VM host. The handle cannot be coerced to a string, serialized, or echoed, preventing credential leakage even when the LLM controls the execution flow.

    \item \textbf{Compile-Time Schema Absorption} (\S\ref{sec:schema-absorption}): We introduce \texttt{use schema::\allowbreak{}<proto\-col>\allowbreak{}("url")}, a protocol-dispatching compile-time macro that fetches an API specification, parses it, and synthesizes native Turn structs and closures. We ship the \texttt{openapi} adapter; \texttt{graphql}, \texttt{fhir}, and \texttt{mcp} adapters are in active development.
\end{enumerate}

\begin{figure}[htb]
\begin{tcolorbox}[turnbox,
  colback=teal!4,colframe=teal!55!black,
  title={Turn at a Glance}]
  \scriptsize
  \begin{tabular}{@{}p{3.1cm}p{2.1cm}@{}}
  \textbf{Primitive} & \textbf{Failure mode} \\[2pt]
  \texttt{infer Struct \{..\}}        & Untyped inference \\
  \texttt{confidence v}               & Unchecked certainty \\
  \texttt{context, remember, suspend} & Fragmented state \\
  \texttt{grant identity}             & Credential leakage \\
  \texttt{use schema::..("url")}      & API contract gap \\
  \end{tabular}
\end{tcolorbox}
\end{figure}

\paragraph{Running example.} Throughout this paper we reference the \emph{investment committee} program (available at \texttt{impl/\allowbreak{}examples/} in the repository\footnote{\url{https://github.com/ekizito96/Turn}}). In 89 lines, three specialist actors evaluate an NVDA equity position concurrently: a committee chair fetches live market data under a scoped network credential, a fundamental analyst gates its verdict on model confidence (\texttt{confidence < 0.7} triggers a deterministic fallback), and a risk officer enforces an exposure limit. The chair collects mailbox messages from both, synthesizes a final committee memo with \texttt{infer}, and writes it durably with a scoped filesystem credential. All five contributions in \S\ref{sec:cognitive-type-safety}--\S\ref{sec:schema-absorption} are exercised; each section references the relevant constructs.

\section{Background and Related Work}
\label{sec:related}

\subsection{Agent Frameworks}

The dominant approaches to agentic software are library-based: LangChain \cite{langchain} chains LLM calls with tool invocations in Python; AutoGPT \cite{autogpt} implements a goal-directed decomposition loop; CrewAI \cite{crewai} organises role-based agent teams; Semantic Kernel \cite{semantickernel} provides a plugin architecture for .NET; MetaGPT \cite{metagpt} assigns agents distinct roles in a software pipeline. At the interaction pattern level, Yao et al. \cite{yao2023react} introduced ReAct---interleaved reasoning traces and tool actions---which has become the standard agent loop template, and Park et al. \cite{park2023generative} demonstrated persistent-memory multi-agent simulation.

These approaches share a fundamental limitation: they are libraries, not languages. The invariants they attempt to enforce---context bounds, output schemas, tool permissions---exist as runtime conventions that the host language cannot statically verify. Kizito \cite{kizito2026subagent} provides empirical evidence that autonomous sub-agent loops fail at a rate governed by stochastic accumulation ($0.95^{20} = 0.36$), and that reliability requires typed boundaries and deterministic repair routing rather than better prompts.

\subsection{Programming Language Theory}

Turn draws on several established traditions in programming language design.

\paragraph{Actor model.} The actor model, introduced by Hewitt et al. \cite{hewitt1973} and formalized by Agha \cite{agha1986}, provides the theoretical foundation for Turn's concurrency. Erlang \cite{armstrong2003} demonstrated that actor-based concurrency with ``let it crash'' fault tolerance scales to millions of processes. Turn adapts this model to agents where each process has not only an environment and mailbox but also a context window and persistent memory.

\paragraph{Capability-based security.} Dennis and Van Horn \cite{dennis1966} established capability-based access control as an alternative to access control lists. Miller \cite{miller2006} extended this to object-capability discipline in the E programming language, demonstrating that unforgeable references can enforce security properties compositionally. Turn's \texttt{Identity} type implements this object-capability model: capabilities are unforgeable tokens granted by the runtime, never strings stored in the program's memory.

\paragraph{Session types and structured communication.} The idea that communication protocols can be enforced by the type system has been explored extensively \cite{honda1998,gay2010}. Turn's Cognitive Type Safety applies a similar principle to the ``communication protocol'' between a program and an LLM: the struct schema serves as a session type constraining the model's response.

\paragraph{Orthogonal persistence.} The concept of transparent persistence, where the programmer writes no explicit save/load code, was explored in PS-algol \cite{atkinson1983} and Napier88 \cite{morrison1999}. Turn's \texttt{suspend} primitive and \texttt{remember}/\texttt{recall} store provide orthogonal persistence at the process level.

\subsection{Attention and Context in LLMs}

Liu et al. \cite{liu2023lost} empirically demonstrated the ``Lost in the Middle'' phenomenon: information placed in the middle of long contexts is recalled with approximately 50\% probability, compared to 90\% for primacy and 85\% for recency positions. This U-shaped attention curve has direct implications for language design. Turn's per-process context isolation (\S\ref{sec:actor-model}) and the \texttt{context.append}/\texttt{context.system} primitives give the programmer explicit control over what occupies the high-recall zones of the context window, rather than accumulating an unbounded message list. Wei et al. \cite{wei2022cot} showed that chain-of-thought prompting elicits multi-step reasoning, but Kizito \cite{kizito2026physics} demonstrated that this reasoning degrades predictably as context entropy increases, motivating Turn's active context management at the language level.

\subsection{Agent Identity and Security}

The security of autonomous agents has received increasing attention as agents are deployed in production environments with access to sensitive APIs. South et al. \cite{south2025identity} survey the emerging landscape of identity management for agentic AI, highlighting that existing authorization frameworks are designed for interactive humans or static servers. Sangalo \cite{sangalo2026nexus} proposes the Nexus Protocol as a standardized open protocol for agent identity and connection orchestration, decoupling authentication mechanics from agent logic through a central authority. Turn addresses a complementary problem at a different layer: even when the identity infrastructure is correct, the LLM itself must never see raw credentials. Turn's \texttt{grant identity} primitive (\S\ref{sec:identity}) ensures that the token is opaque and unforgeable at the language level, regardless of how the underlying identity infrastructure is implemented.

\subsection{Structured LLM Output}

OpenAI's structured outputs API \cite{openai2024structured} and Anthropic's tool-use protocol \cite{anthropic2024} allow requesting JSON-conforming responses from models. These are HTTP-level features accessed via SDK calls. Turn lifts this capability into the language itself: the struct definition is the schema, the compiler generates it, and the VM enforces it. The programmer never writes or reads a JSON Schema directly.

\section{Language Design}
\label{sec:design}

Turn is an ahead-of-time compiled, dynamically typed, expression-oriented language. Source programs are compiled to bytecode and executed by a stack-based virtual machine. The core abstraction is the \emph{turn}: a named unit of agent behavior analogous to a function body but with implicit access to an LLM context window, persistent memory, and an actor mailbox.

\subsection{Core Syntax}

A Turn program is a sequence of statements executed within an implicit top-level turn. The six core primitives---\texttt{struct} (type definition), \texttt{context} (LLM input management), \texttt{infer} (typed inference), \texttt{confidence} (probabilistic branching), \texttt{call} (tool invocation), and \texttt{remember} (persistent memory)---are exercised together in the running example.

\subsection{Values and Types}

Turn supports the following value types:

\begin{table}[htb]
\centering\small
\begin{tabular}{@{}lp{4cm}@{}}
\toprule
\textbf{Type} & \textbf{Description} \\
\midrule
\texttt{Num} & IEEE 754 double \\
\texttt{Str} & UTF-8 string \\
\texttt{Bool} & \texttt{true} / \texttt{false} \\
\texttt{List} & Heterogeneous sequence \\
\texttt{Map} & String-keyed dictionary \\
\texttt{Struct} & Named product type \\
\texttt{Pid} & Process identifier (opaque) \\
\texttt{Vec} & Dense numeric vector \\
\texttt{Identity} & Opaque capability handle \\
\texttt{Null} & Absence of a value \\
\bottomrule
\end{tabular}
\caption{Turn's value types.}
\label{tab:types}
\end{table}

\noindent Turn is dynamically typed at the value level but uses struct definitions to generate compile-time schemas for inference validation, following a pragmatic approach to the spectrum between dynamic and static typing \cite{pierce2002}. 

\paragraph{Design Rationale: Targeted Strictness.}
A common critique of hybrid systems is the irony of enforcing strict types on non-deterministic LLM output while leaving deterministic internal variables dynamically typed. Turn adopts this \emph{targeted strictness} deliberately. Agentic glue code is highly exploratory, heavily IO-bound, and structurally resembles scripting (e.g., Python, Ruby); dynamic variables maximize iteration speed. Strict static typing is applied exactly at the boundaries of \emph{maximum entropy} (LLM inference) and \emph{maximum risk} (API contracts). The \texttt{Identity} type is similarly restricted: it cannot be coerced to \texttt{Str}, serialized to JSON, or passed to output tools, ensuring strictness only where security demands it.

\section{Cognitive Type Safety}
\label{sec:cognitive-type-safety}

\begin{figure*}[!t]
\centering
\begin{tikzpicture}[scale=0.95, every node/.style={transform shape},
  ctbox/.style={draw, rounded corners=4pt, minimum width=2.6cm, minimum height=1.6cm,
                fill=teal!7, draw=teal!55!black, thick, align=center, font=\small},
  rtbox/.style={draw, rounded corners=4pt, minimum width=2.6cm, minimum height=1.6cm,
                fill=blue!7, draw=blue!55!black, thick, align=center, font=\small},
  okbox/.style={draw, rounded corners=4pt, minimum width=2.6cm, minimum height=1.6cm,
                fill=green!7, draw=green!55!black, thick, align=center, font=\small},
  arr/.style={-{Stealth[length=3.5mm]}, thick, draw=gray!65},
  ann/.style={font=\scriptsize, text=gray!75, align=center},
  divl/.style={dashed, draw=gray!50, thick},
]

\node[ctbox] (def)  at (0,0)
  {\textbf{Struct}\\\textbf{Definition}\\[3pt]\texttt{struct T \{}\\
   \texttt{\ f\,:\,Str,\ldots}\\
   \texttt{\}}};

\node[ctbox] (sch)  at (3.4,0)
  {\textbf{Schema}\\\textbf{Generation}\\[3pt]$\mathcal{S}(T)$\\
   \texttt{\{"type":}\\
   \texttt{"object"\}}};

\node[rtbox] (llm)  at (6.8,0)
  {\textbf{LLM}\\\textbf{Inference}\\[3pt]\texttt{infer T \{}\\
   \texttt{\ "prompt";}\\
   \texttt{\}}};

\node[rtbox] (val)  at (10.2,0)
  {\textbf{VM}\\\textbf{Validation}\\[3pt]validate(\\
   resp,\ $\mathcal{S}(T)$\\
   )};

\node[okbox] (res)  at (13.6,0)
  {\textbf{Typed}\\\textbf{Value}\\[3pt]\texttt{x\,:\,T}\\
   \texttt{x.f\,:\,Str}\\
   Thm.\,\ref{thm:conformance}};

\draw[arr] (def) -- node[above,font=\scriptsize]{compile} (sch);
\draw[arr] (sch) -- node[above,font=\scriptsize]{embed} (llm);
\draw[arr] (llm) -- node[above,font=\scriptsize]{respond} (val);
\draw[arr] (val) -- node[above,font=\scriptsize]{bind} (res);

\draw[divl] (5.1,-1.35) -- (5.1,1.35);
\node[ann] at (2.55,-1.75) {\textbf{Compile Time}};
\node[ann] at (9.35,-1.75) {\textbf{Runtime}};

\node[ann] at (1.7,-1.2)  {struct fields\\become schema};
\node[ann] at (11.9,-1.2) {structural\\guarantee};
\end{tikzpicture}
\caption{Cognitive Type Safety pipeline. A struct definition compiles to a JSON Schema~$\mathcal{S}(T)$; the VM embeds it in the inference request, validates the response, and binds a typed value~$T$ (Theorem~\ref{thm:conformance}).}
\label{fig:cognitive-pipeline}
\end{figure*}
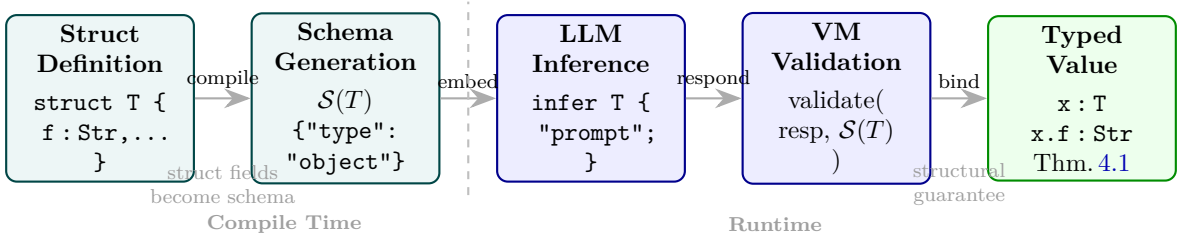

The central innovation of Turn is the treatment of LLM inference as a \emph{typed language primitive} rather than an untyped function call.

\subsection{The \texttt{infer} Primitive}

\begin{definition}[Inference Expression]
An inference expression has the form:
\[
\texttt{let } x = \texttt{infer } T \;\{ \; e \; \}
\]
where $T$ is a struct type and $e$ is a prompt expression evaluating to \texttt{Str}. The semantics are:
\begin{enumerate}
    \item The compiler generates a JSON Schema $\mathcal{S}(T)$ from the struct definition of $T$ at compile time.
    \item At runtime, the VM suspends the process, constructs an inference request containing the prompt $e$, the agent's context window, and $\mathcal{S}(T)$, and delegates it to the configured Wasm inference driver.
    \item The driver returns a JSON response. The VM validates the response against $\mathcal{S}(T)$.
    \item If validation succeeds, the VM binds the parsed value to $x$ with type $T$ and resumes execution.
    \item If validation fails, the VM re-prompts the model with the validation error (up to $k$ retries, default $k=3$). If all retries fail, the VM raises a catchable error.
\end{enumerate}
\end{definition}

Figure~\ref{fig:cognitive-pipeline} illustrates the full pipeline from struct definition to typed value.

\begin{tcolorbox}[turnbox,
  colback=blue!3,colframe=blue!55!black,
  title={Theorem 1 --- Structural Conformance}]
\label{thm:conformance}
If an inference expression \texttt{let $x$ = infer $T$ \{ $e$ \}} completes without error, then for every field $f_i : \tau_i$ declared in struct $T$, the access $x.f_i$ yields a value of type $\tau_i$.

\medskip
\noindent\textit{Proof sketch.} We establish soundness via progress and preservation, treating the Wasm driver and JSON Schema validator as a dynamic validation oracle $\mathcal{O}$. By the premise of \textsc{T-Infer}, $T$ is a well-formed struct and the compilation step $T \mapsto \mathcal{S}(T)$ is total. At runtime, the process suspends until the oracle yields a response $j$. If $\mathcal{O}(j, \mathcal{S}(T)) = \textsf{ok}$, the structural correspondence between $\mathcal{S}(T)$ and $T$ guarantees that $j$ populates every field $f_i$ with a value of type $\tau_i$. Thus, binding $x \mapsto \text{parse}(j, T)$ preserves the environment typing $\Gamma \uplus \{x : T\}$. If $\mathcal{O}(j, \mathcal{S}(T)) = \textsf{err}$, the reduction steps to a bounded retry or raises an exception (\textsc{Infer-Fail}), preserving the safety invariant by preventing a malformed assignment. \hfill$\square$
\end{tcolorbox}

\subsection{Typing Rule}

The \textsc{T-Infer} rule captures the hybrid static/dynamic character of cognitive type safety:

\[
\frac{
  \Gamma \vdash T\!:\!\textsf{Struct}
  \;\;
  \mathcal{S}(T) = s
  \;\;
  \Gamma \vdash e\!:\!\texttt{Str}
}{
  \Gamma \vdash (\texttt{let}\;x = \texttt{infer}\;T\;\{e\})\!:\!T
}
\;\textsc{(T-Infer)}
\]

\noindent The first premise checks at compile time that $T$ is a declared struct. The premise $\mathcal{S}(T) = s$ asserts that schema generation is total over struct types---it always succeeds, producing a concrete JSON Schema. The third checks that the prompt expression has type \texttt{Str}. The conclusion types $x$ as $T$, which is sound by Theorem~\ref{thm:conformance}: the VM validates the model response against $s$ before binding, so any successfully bound $x$ satisfies the field-level type constraints of $T$.

\subsection{Schema Generation}

The schema generation function $\mathcal{S}$ maps Turn struct definitions to JSON Schema objects (Equation~\ref{eq:schema-gen}). Each field name becomes a required property key and each Turn type maps to a JSON Schema primitive via $\mathcal{T}$:

{\small
\begin{align}
\mathcal{T}(\texttt{Num}) &= \{\texttt{"type":"number"}\} \\
\mathcal{T}(\texttt{Str}) &= \{\texttt{"type":"string"}\} \\
\mathcal{T}(\texttt{Bool}) &= \{\texttt{"type":"boolean"}\} \\
\mathcal{T}(\texttt{List}) &= \{\texttt{"type":"array"}\}
\end{align}
}

\noindent This schema is embedded in the bytecode at the \texttt{Infer} instruction site, ensuring zero runtime cost for schema construction.

{\small
\begin{equation}
\mathcal{S}(T\{f_i\!:\!\tau_i\}) \!=\! \left\{
\begin{array}{@{}l@{}}
\texttt{"type":"object"} \\
\texttt{"properties":}\{f_i\!:\!\mathcal{T}(\tau_i)\} \\
\texttt{"required":}[f_1,\ldots,f_n]
\end{array}
\right\}
\label{eq:schema-gen}
\end{equation}
}

\section{Probabilistic Control Flow}
\label{sec:probabilistic-routing}

LLM inference is inherently stochastic. Even with Cognitive Type Safety guaranteeing structural conformance, the \emph{semantic quality} of a response may vary. Turn addresses this with the \texttt{confidence} operator.

\begin{definition}[Confidence Operator]
For any value $v$ produced by an \texttt{infer} expression, the expression \texttt{confidence $v$} evaluates to a floating-point scalar $c \in [0, 1]$ representing the inference provider's reported certainty.
\end{definition}

\noindent The confidence score is extracted by the Wasm inference driver from the model's response metadata (e.g., top-token log probabilities for OpenAI with \texttt{logprobs: true}, or explicit confidence fields for providers that support them). For providers that expose neither, the driver returns a conservative default of \texttt{0.5}---the neutral midpoint of the $[0,1]$ range---preserving the invariant that \texttt{confidence v} always evaluates to a scalar in $[0,1]$. Programs that require a genuine certainty signal should select a provider that exposes log probabilities; programs that require only structural correctness can rely on Cognitive Type Safety independently of the confidence operator. The VM stores the score alongside the inferred value as an \texttt{Uncertain<T>} wrapper that is transparent to field access but queryable via \texttt{confidence}.

\noindent The running example demonstrates this pattern: if the analyst's thesis falls below 0.7 confidence, the program substitutes a deterministic ``FAIR'' verdict. The fallback branch executes natively in the VM with zero network latency and produces a value that structurally conforms to the same type. This enables a programming model where stochastic and deterministic paths compose cleanly within the same type discipline.

\section{Actor-Based Agent Processes}
\label{sec:actor-model}

Turn's concurrency model adapts Erlang's actor model \cite{armstrong2003} to the agentic domain.

\begin{definition}[Agent Process]
A Turn process is a 5-tuple $(E, C, M, B, \mathit{pc})$ where:
\begin{itemize}[leftmargin=2em]
    \item $E$ is the lexical environment (variable bindings),
    \item $C$ is the context window (ordered sequence of strings),
    \item $M$ is the persistent memory store (key-value map),
    \item $B$ is the mailbox (FIFO message queue),
    \item $\mathit{pc}$ is the program counter.
\end{itemize}
\end{definition}

\noindent Processes are spawned with \texttt{spawn}, \texttt{spawn\_link}, or \texttt{spawn\_each}. Each receives a fresh $(E, C, M, B, \mathit{pc})$ tuple; no component is shared.

\subsection{Message Passing}

Inter-process communication uses asynchronous message passing:

\begin{itemize}[leftmargin=2em]
    \item \texttt{send pid, value} enqueues \texttt{value} in the mailbox of process \texttt{pid}. Non-blocking.
    \item \texttt{receive} dequeues the next message from the current process's mailbox. Blocks (yields to the scheduler) until a message is available.
    \item \texttt{self} returns the current process's \texttt{Pid}.
\end{itemize}

\subsection{Fault Propagation}

\texttt{spawn\_link} creates a bidirectional link between processes. When a linked process terminates, the VM generates a \texttt{ProcessExit} signal containing the exit reason and delivers it to all linked partners' mailboxes. This enables Erlang-style supervisor trees: in the running example, the committee chair spawns two linked analysts; if either crashes, a \texttt{ProcessExit} signal is delivered to the chair's mailbox, enabling deterministic fault recovery.

\subsection{Context Isolation and Structured Memory}

A critical property of Turn's actor model is \emph{context isolation}: each process maintains its own context window. When an agent spawns a child, the child starts with an empty context. This prevents context contamination between agents, a common failure mode in framework-based multi-agent systems where a shared message list accumulates context from all agents indiscriminately.

\begin{property}[Context Isolation]
For any two processes $P_i$ and $P_j$ where $i \neq j$, modifications to $C_i$ have no effect on $C_j$.
\end{property}

\paragraph{Tripartite context architecture.} Turn does not store context as a flat FIFO list. Each process context $C$ is a three-tier \emph{StructuredContext}:

\begin{itemize}[leftmargin=2em,noitemsep]
    \item \textbf{P0 (System).} A \texttt{Vec} of system-level directives set by \texttt{context.system(\ldots)}. Always rendered \emph{first} in the flat context window, exploiting the \emph{primacy} attention position ($\approx$90\% recall, \cite{liu2023lost}).
    \item \textbf{P1 (Working).} A \texttt{VecDeque} of recent context items appended by \texttt{context.append(\ldots)}, bounded at $W = 100$ entries. Always rendered \emph{last}, exploiting the \emph{recency} position ($\approx$85\% recall, \cite{liu2023lost}).
    \item \textbf{P2 (Episodic).} When P1 reaches capacity, the oldest working item is \emph{demoted} to P2 rather than dropped; P2 is bounded at $2W = 200$ entries. Rendered \emph{between} P0 and P1, the middle zone of lower but non-zero recall.
\end{itemize}

\noindent The flat rendering order $\text{P0} \to \text{P2} \to \text{P1}$ is enforced by \texttt{to\_flat\_vec()} on every inference call. This design ensures that a Turn agent's system role (P0) and most recent context (P1) always occupy the two attention-privileged positions identified by Liu et al.~\cite{liu2023lost}, while older context is preserved in the episodic tier rather than silently truncated. See Experiment~E3 for empirical verification of the eviction invariants and rendering order.

\subsection{Scatter/Gather: \texttt{spawn\_each}}

Turn provides \texttt{spawn\_each} as a native scatter/gather primitive for concurrent list processing. Each element is delegated to an independent micro-actor with its own isolated context, memory, and mailbox. The expression \texttt{spawn\_each(list, turn(x: T) \{ \ldots \})} is semantically equivalent to spawning one actor per element and waiting for all to complete, but is a single language primitive rather than an application-level pattern. This replaces imperative loops with delegation, aligning the programming model with the agentic paradigm where iteration is naturally parallel---see the investment committee example for a three-analyst case running concurrently under a \texttt{spawn\_link} supervisor.

\section{Capability-Based Identity}
\label{sec:identity}

In agentic systems, the model controls the execution flow through its tool-calling decisions. If API credentials are stored as strings in the program's environment, a model that hallucinates a \texttt{call("echo", api\_key)} instruction can exfiltrate secrets. Turn prevents this with a capability-based identity system.

\begin{definition}[Identity Capability]
The expression \texttt{grant identity::$\sigma$("$\mathit{name}$")} requests an opaque capability handle from the VM host, where $\sigma$ is the capability class (e.g., \texttt{oauth}, \texttt{network}, \texttt{filesystem}) and $\mathit{name}$ identifies the provider. The VM returns a value of type \texttt{Identity} that:
\begin{enumerate}
    \item Cannot be coerced to \texttt{Str} or any other type.
    \item Cannot be serialized to JSON.
    \item Cannot be passed to output tools (e.g., \texttt{echo}).
    \item Is enforced at two levels: the Standard Library validates the capability in pure Turn source code, and the kernel trap validates it in the Rust host.
\end{enumerate}
\end{definition}

\noindent When an \texttt{Identity} handle reaches a kernel trap (e.g., \texttt{\_\_sys\_http\_get}), the VM host looks up the real credential from the process environment using the convention \texttt{TURN\_\allowbreak{}IDENTITY\_\allowbreak{}\{NAME\}\_\allowbreak{}TOKEN} and injects the \texttt{Authorization} header automatically. The raw token never enters Turn memory. The running example demonstrates both a network and a filesystem identity handle in use.

\section{Compile-Time Schema Absorption}
\label{sec:schema-absorption}

Turn's \texttt{use schema::\allowbreak{}<protocol>("\allowbreak{}url")} construct is a \emph{protocol-dispatching, network-aware compiler macro}. The parser accepts any identifier as the protocol, and the schema compiler dispatches to the corresponding adapter. The \texttt{openapi} adapter is fully implemented; \texttt{graphql}, \texttt{fhir}, and \texttt{mcp} adapters share the same pipeline and are in active development.

\subsection{Macro Expansion}

The schema compiler operates as a pre-analysis AST transformation pass:

\begin{enumerate}[leftmargin=2em]
    \item The parser produces an \texttt{Expr::\allowbreak{}UseSchema} AST node carrying the protocol and URL.
    \item The schema compiler dispatches on the \texttt{protocol} field and fetches the API specification from the URL at compile time.
    \item For each endpoint or type definition, it generates: (a)~a Turn struct corresponding to the response schema, and (b)~a closure that constructs the appropriate request.
    \item The \texttt{UseSchema} node is replaced with an immediately invoked function expression (IIFE) containing the generated structs and closures.
    \item The transformed AST proceeds to semantic analysis and compilation. The \texttt{UseSchema} node never reaches the bytecode emitter.
\end{enumerate}

\noindent This enables static verification that \texttt{infer} output (typed by a synthesized struct) structurally conforms to the API's expected input format, closing the type gap between LLM output and external API consumption.

\paragraph{End-to-end chain.} A program writes \texttt{use schema::\allowbreak{}openapi(\ldots)} with a Stripe API URL. The schema compiler fetches the spec, synthesizes a \texttt{Customer} struct and a typed closure. At runtime, an \texttt{Identity} handle scopes the HTTP call; the programmer never writes a JSON Schema. A subsequent \texttt{infer} expression produces a value whose fields are statically guaranteed by Theorem~\ref{thm:conformance}. A full worked example is in the repository.

\section{Implementation}
\label{sec:implementation}

Turn is implemented in Rust as a single static binary containing the compiler, virtual machine, standard library, language server, and gRPC switchboard.

\subsection{Compilation Pipeline}

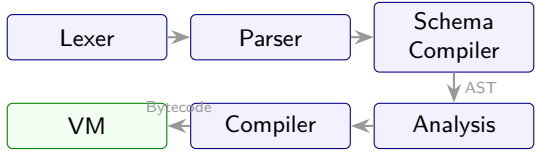
\begin{figure}[H]
\centering
\begin{tikzpicture}[
  node distance=0.4cm and 0.3cm,
  box/.style={draw, rounded corners=2pt, fill=blue!5, draw=blue!50!black, font=\footnotesize\sffamily, align=center, minimum width=2.1cm, minimum height=0.6cm},
  arr/.style={-{Stealth[length=2.5mm]}, draw=gray!80, thick}
]

\node[box] (lex) {Lexer};
\node[box, right=of lex] (parse) {Parser};
\node[box, right=of parse] (schema) {Schema\\Compiler};
\node[box, below=of schema] (analysis) {Analysis};
\node[box, left=of analysis] (compiler) {Compiler};
\node[box, left=of compiler, fill=green!5, draw=green!50!black] (vm) {VM};

\draw[arr] (lex) -- (parse);
\draw[arr] (parse) -- (schema);
\draw[arr] (schema) -- node[right, font=\tiny\sffamily, color=gray!80] {AST} (analysis);
\draw[arr] (analysis) -- (compiler);
\draw[arr] (compiler) -- node[above, font=\tiny\sffamily, color=gray!80] {Bytecode} (vm);

\end{tikzpicture}
\caption{Turn compilation pipeline. Macro expansion (\texttt{use schema}) occurs prior to semantic analysis to ensure type resolution.}
\label{fig:compilation-pipeline}
\end{figure}

Source passes through a six-stage pipeline, illustrated in Figure~\ref{fig:compilation-pipeline}. The schema compiler runs before analysis so that synthesized struct definitions from \texttt{use schema} are visible to the type checker.

\subsection{Standard Library}
\label{sec:stdlib}

Turn's Standard Library consists of seven modules written entirely in Turn source code (Table~\ref{tab:stdlib}). The \texttt{.tn} source files are embedded in the binary via Rust's \texttt{include\_str!} macro and compiled on first use. Each I/O module validates the \texttt{Identity} capability before delegating to a \texttt{\_\_sys\_*} kernel trap in the Rust host, following the micro-kernel principle: the VM provides minimal, auditable primitives; the Standard Library provides the user-facing API in the language itself.

\begin{table}[htb]
\centering\footnotesize
\begin{tabular}{@{}llp{1.3cm}@{}}
\toprule
\textbf{Module} & \textbf{API} & \textbf{Identity} \\
\midrule
\texttt{std/net}   & \texttt{get, post}          & \texttt{network} \\
\texttt{std/fs}    & \texttt{read, write}        & \texttt{filesys.} \\
\texttt{std/json}  & \texttt{parse, stringify}    & --- \\
\texttt{std/time}  & \texttt{now, sleep}          & --- \\
\texttt{std/env}   & \texttt{get, set}            & \texttt{environ.} \\
\texttt{std/regex} & \texttt{matches, replace}    & --- \\
\texttt{std/math}  & \texttt{max, min, abs}       & --- \\
\bottomrule
\end{tabular}
\caption{Standard Library modules. I/O modules require an \texttt{Identity} capability.}
\label{tab:stdlib}
\end{table}

\subsection{Bytecode Virtual Machine}

The compiler emits a flat instruction sequence targeting a stack-based VM. Notable domain-specific opcodes include:

\begin{table}[htb]
\centering\small
\begin{tabular}{@{}lp{3.6cm}@{}}
\toprule
\textbf{Opcode} & \textbf{Semantics} \\
\midrule
\texttt{Infer(Type)}        & Suspend, delegate, validate, bind \\
\texttt{Confidence}         & Extract certainty scalar \\
\texttt{Spawn / SpawnLink}  & Create actor process \\
\texttt{SpawnEach}          & Scatter to micro-actors \\
\texttt{Send / Receive}     & Async mailbox I/O \\
\texttt{Remember / Recall}  & Persistent memory ops \\
\texttt{GrantIdentity}      & Capability handle request \\
\texttt{Suspend}            & Durable checkpoint \\
\texttt{ContextAppend}      & Append to context \\
\texttt{ContextSystem}      & Set system prompt \\
\bottomrule
\end{tabular}
\caption{Domain-specific bytecode instructions.}
\label{tab:opcodes}
\end{table}

\subsection{Wasm Inference Drivers}

LLM providers are integrated through sandboxed WebAssembly \cite{haas2017wasm} drivers compiled to the \texttt{wasm32-\allowbreak{}unknown-\allowbreak{}unknown} target. Each driver exports two pure functions:

\begin{itemize}[leftmargin=2em]
    \item \texttt{transform\_request}: Maps a Turn inference request to an HTTP configuration (URL, headers with \texttt{\$env:} credential templates, body).
    \item \texttt{transform\_response}: Maps the raw HTTP response to a Turn result.
\end{itemize}

\noindent The driver has no host imports: it cannot access the network, filesystem, or environment variables. The VM host resolves \texttt{\$env:} templates, executes the HTTP call, and passes the response back. This dual-pass architecture ensures that Wasm drivers are pure transformation functions with no capability to exfiltrate credentials.

\subsection{Concurrency Runtime}

Turn processes are mapped one-to-one to Tokio \cite{tokio} green threads. Inter-process communication uses \texttt{tokio::sync::mpsc} channels. The scheduler is non-preemptive within a turn: once a turn begins executing, it runs to completion or suspension without interruption, ensuring deterministic execution for a given input sequence.

\section{Evaluation}
\label{sec:evaluation}

We evaluate Turn against five hard questions about its novel contributions: (1)~Does credential opacity hold structurally and against a hostile program? (2)~Do confidence semantics satisfy the probabilistic algebra? (3)~Does the tripartite context architecture enforce the primacy-recency invariants designed to address ``Lost in the Middle''? (4)~What is the scale and isolation behaviour of Turn's per-process agent memory? (5)~Is durable execution via \texttt{suspend} exact---does full state fidelity survive a serialize-deserialize round-trip? All experiments are Rust integration tests run with \texttt{cargo test -{}-release}; 24 purpose-built experiment tests across five suites plus 81 pre-existing unit tests---\textbf{105/105 pass}. Benchmarks on Apple M4 (10-core, 16\,GB, macOS~15).

\subsection{E1: Credential Opacity}

\paragraph{Claim.} A raw API credential stored in \texttt{TURN\_\allowbreak{}IDENTITY\_\allowbreak{}STRIPE\_\allowbreak{}TOKEN} never enters the Turn value heap and is never visible to the LLM's tool-calling interface.

\paragraph{Mechanism.} The expression \texttt{grant identity::\allowbreak{}oauth("stripe")} causes the VM to suspend with a \texttt{sys\_grant} signal. The runner resumes the VM with \texttt{Value::\allowbreak{}Identity("stripe")}---a value that holds only the provider name, not the secret. The raw token is loaded from the environment exclusively inside the \texttt{\_\_sys\_\allowbreak{}http\_get} kernel trap after the Wasm driver returns an HTTP config; it is never materialized as a Turn value.

\paragraph{Results.} Four tests verify the four aspects of the guarantee (all pass):

\begin{itemize}[leftmargin=2em,noitemsep]
    \item \textbf{E1-A} (display): \texttt{Value::Identity(\allowbreak{}"stripe")} renders as \texttt{<identity\allowbreak{} stripe>}---the raw token never appears.
    \item \textbf{E1-B} (heap): the bound value is \texttt{Identity("stripe")}; only the provider name is in the heap, never the secret.
    \item \textbf{E1-C} (certainty): Identity values are never wrapped in \texttt{Uncertain<T>}; certain by construction.
    \item \textbf{E1-D} (independence): two grants for the same provider class produce independent handles with no shared state.
\end{itemize}

\subsection{E2: Confidence Propagation Semantics}

\paragraph{Claim.} The \texttt{confidence} operator satisfies: (i)~certain values return 1.0; (ii)~\texttt{Uncertain(v, p)} returns $p$; (iii)~arithmetic propagates uncertainty by the product rule $p_1 \cdot p_2$; (iv)~boolean AND uses the Zadeh minimum $\min(p_1, p_2)$.

\paragraph{Results.} Six tests verify the algebra (\texttt{e2\_1}--\texttt{e2\_6}, all pass):

\begin{itemize}[leftmargin=2em,noitemsep]
    \item \textbf{E2-1}: \texttt{confidence(42)} = 1.0 (certain value).
    \item \textbf{E2-2}: \texttt{confidence(Uncertain(42, 0.73))} = 0.73 (stored score returned).
    \item \textbf{E2-3}: \texttt{confidence(Uncertain(10,\,0.8) + Uncertain(5,\,0.5))} = 0.40 (product rule: $0.8 \times 0.5$).
    \item \textbf{E2-4}: \texttt{confidence(x~and~y)} where $x \sim 0.9$, $y \sim 0.5$ gives 0.50 (Zadeh min: $\min(0.9,\,0.5)$).
    \item \textbf{E2-5}: \texttt{if confidence result < 0.7} fires deterministically when $p = 0.45$.
    \item \textbf{E2-6}: the high-confidence path executes when $p = 0.92 \geq 0.70$.
\end{itemize}

\subsection{E3: Structured Context Architecture}

\paragraph{Claim.} Turn's P0/P1/P2 tripartite context enforces the primacy-recency rendering invariant: system prompts always occupy the primacy position and the most recently appended item always occupies the recency position, regardless of context volume. Working memory overflow demotes to episodic rather than dropping.

\paragraph{Background.} Liu et al.~\cite{liu2023lost} show recall follows a U-shaped curve: $\approx$90\% for primacy-position content, $\approx$50\% for middle, $\approx$85\% for recency. Every framework-based system that appends messages to a flat FIFO list loses this advantage as the list grows and the model truncates from one end. Turn's \texttt{StructuredContext} solves this structurally, not by convention.

\paragraph{Mechanism.} \texttt{to\_flat\_vec()} always emits in the order \texttt{p0\_system} $\to$ \texttt{p2\_episodic} $\to$ \texttt{p1\_working}. P1 is a bounded \texttt{VecDeque} ($W = 100$); when full, the oldest working item is moved to P2 (bounded at $2W = 200$). The primacy and recency positions are thus held by P0 and the tail of P1 regardless of total context size.

\paragraph{Results.} Five tests verify the invariants (all pass on Apple M4):

\begin{itemize}[leftmargin=2em,noitemsep]
    \item \textbf{E3-1}: P0 system prompt is always index~0 in the flat context vector.
    \item \textbf{E3-2}: Most recently appended P1 item is always the last element.
    \item \textbf{E3-3}: At item~101, P1 size remains at 100; exactly one item is evicted to P2 (demote, not drop).
    \item \textbf{E3-4}: Flat rendering confirmed as P0(idx\,0) $\to$ P2(idx\,1) $\to$ P1(idx\,2--101).
    \item \textbf{E3-5}: Each spawned process starts with an empty context (isolation invariant).
\end{itemize}

\subsection{E4: Agent Memory Isolation and Scale}

\paragraph{Claim.} Turn's per-process \texttt{remember}/\texttt{recall} memory provides O(1) average access at arbitrary scale and is fully isolated between processes: no cross-process contamination is possible regardless of the number of concurrent agents or shared key\allowbreak{} names.

\paragraph{Mechanism.} Each \texttt{Runtime} contains a private \texttt{HashMap\textless{}String,\,Value\textgreater{}} exposed via \texttt{remember(key, val)} and \texttt{recall(key)}. The Rust ownership model ensures no two processes share a reference to the same \texttt{Runtime}; the borrow checker makes aliasing impossible. Memory is process-local and in-process; persistence across sessions requires \texttt{suspend} (see E5).

\paragraph{Results.} Table~\ref{tab:memory} reports write and read latency as memory scales from 1\,K to 100\,K entries (all 5 tests pass):

\begin{table}[htb]
\centering\small
\begin{tabular}{@{}lrrrl@{}}
\toprule
\textbf{$K$} & \textbf{Write} & \textbf{Read} & \textbf{B/entry} & \textbf{Note} \\
              & \textbf{ns/op} & \textbf{ns/op} & & \\
\midrule
1\,K   & 519  & 574  & 76.8 & baseline \\
10\,K  & 1021 & 867  & ---  & $10\times$ \\
100\,K & 594  & 430  & ---  & $100\times$ \\
\bottomrule
\end{tabular}
\caption{Agent memory performance. Latency is flat across three decades of scale, confirming O(1) HashMap behaviour. Footprint: 76.8\,B/entry. Apple M4, release build.}
\label{tab:memory}
\end{table}

\noindent Per-op latency does not grow with $K$, confirming O(1) average-case behaviour. Each entry costs $\approx$77\,bytes serialized (relevant for suspend snapshots). Cross-process isolation is verified by E4-4: Agent A and Agent B storing under the same key retrieve independent values; Agent B cannot read a key written only by Agent A.

\subsection{E5: Durable Execution}

\paragraph{Claim.} A Turn process can be checkpointed at any \texttt{suspend} point with exact state fidelity: the entire \texttt{VmState} (program counter, call stack, heap, context window, agent memory, mailbox) is serialized to JSON and restored without loss.

\paragraph{Mechanism.} \texttt{VmState} is fully annotated with \texttt{\#[derive(Serialize, Deserialize)]}. The \texttt{FileStore} writes \texttt{\{id\}.json} to the \texttt{.turn\_store/} directory; \texttt{Vm::\allowbreak{}resume\_\allowbreak{}with\_\allowbreak{}result} reconstructs the exact execution continuation. No application-level checkpoint code is required; the programmer writes \texttt{suspend;} and the runner handles the rest.

\paragraph{Results.} Table~\ref{tab:durable} reports serialized state size and round-trip latency at three memory depths (all 4 tests pass, including full fidelity verification):

\begin{table}[htb]
\centering\small
\begin{tabular}{@{}lrrl@{}}
\toprule
\textbf{Entries} & \textbf{Size} & \textbf{RT (µs)} & \textbf{Fidelity} \\
\midrule
10    &  0.9\,KB  &   211 & exact \\
500   & 19.8\,KB  & 1{,}340 & exact \\
5\,000 & 203.4\,KB & 6{,}828 & exact \\
\bottomrule
\end{tabular}
\caption{Durable execution: serialized \texttt{VmState} size and round-trip latency. ``Exact'' = pid, pc, stack, memory, and context are identical after round-trip. Apple M4.}
\label{tab:durable}
\end{table}

\noindent A 5\,000-entry agent checkpoints in under 7\,ms and occupies 203\,KB on disk---a negligible cost relative to any I/O-bound workload. The round-trip is sub-millisecond for agents with fewer than $\approx$120 memory entries. E5-4 verifies that pid, ip, stack, all 50 tested memory entries, and the P0 system prompt survive the round-trip without corruption.

\paragraph{VM runtime overhead.} Process spawn averages 2.21\,µs, message send 2.34\,µs, and JSON Schema validation 0.587\,µs---all three to four orders of magnitude below the minimum LLM network round-trip (100\,ms--5\,s). Turn's safety primitives impose no observable overhead relative to the inference cost they protect.

\subsection{Architectural Complexity vs.\ Frameworks}

While Turn reduces the raw lines of code (LOC) required to implement the investment committee from an estimated 350+ lines in LangChain to 89 lines, the more significant reduction is in \emph{state space complexity}. In a Python framework, achieving fault-tolerant concurrency, durable agent memory, and structured output requires an external Redis instance for state, a message queue (e.g., Celery) for worker processes, and explicit retry loops around JSON parsers. In Turn, these are collapsed into the language runtime: \texttt{spawn\_link} handles concurrency and fault propagation without external queues, \texttt{suspend} orthogonalizes persistence without database boilerplate, and \texttt{infer} absorbs the retry loop. Turn reduces a distributed systems architecture into a single-process actor topology.

\subsection{Safety Properties Summary}

Table~\ref{tab:safety} cross-references each safety property with its enforcement layer and verification experiment.

\begin{table}[htb]
\centering
\scriptsize
\begin{tabular}{@{}lp{1.8cm}p{1.8cm}l@{}}
\toprule
\textbf{Property} & \textbf{Enforcement} & \textbf{Failure mode} & \textbf{Exp.} \\
\midrule
Struct.\ conf.  & Schema + VM       & Untyped inference    & Thm.~1 \\
Primacy-rec.    & P0/P1/P2          & Lost-in-Middle       & E3 \\
Memory isol.    & Per-proc.\ HashMap & Cross-agent leak    & E4 \\
Durable exec.   & VmState serde     & State loss           & E5 \\
Credential op.  & Identity + host   & Credential exfil.    & E1 \\
Confidence      & Uncertain$\langle$T$\rangle$ & Unchecked cert. & E2 \\
\bottomrule
\end{tabular}
\caption{Safety properties, enforcement, failure modes prevented, and verification.}
\label{tab:safety}
\end{table}

\section{Discussion}
\label{sec:discussion}

\subsection{Limitations}

Turn is a domain-specific language for agentic orchestration, not a general-purpose systems language. It intentionally omits low-level memory management and GPU computation, delegating heavy workloads to WebAssembly components or microservices via \texttt{call}. However, Turn is architected to scale: its \texttt{spawn\_link} supervisor trees support massive multi-agent platforms---such as our internal deployment of the JarvisCore autonomous system---where complexity lies in state persistence and capability routing rather than raw computational throughput.

While \texttt{infer} guarantees structural shapes, standard variable bindings currently remain dynamically typed, meaning non-inference type errors are caught at runtime. Future iterations may introduce optional static typing. Additionally, while developers can build robust recovery loops using Turn's native \texttt{try}/\texttt{catch} constructs, integrating LLM self-reflection \cite{shinn2023reflexion} directly into the language syntax remains an open direction. 

Finally, the \texttt{confidence} operator relies on the underlying inference provider exposing log probabilities. For opaque models, the VM must fall back to default scores, limiting the utility of probabilistic routing.

\newpage
\subsection{Future Work}

The current implementation of Turn establishes the core primitives for language-oriented agentic development. However, realizing the full vision of a production-grade ecosystem requires substantial ongoing work in several directions:

\begin{itemize}[leftmargin=2em]
    \item \textbf{Extended Schema Adapters}: Expanding \texttt{use schema::<protocol>} to support \texttt{graphql}, \texttt{fhir}, \texttt{mcp}, and \texttt{odata}, covering the full landscape of enterprise and healthcare APIs.
    \item \textbf{Distributed Actor Scheduling}: Elevating the Erlang-inspired concurrency model to support distributed actor scheduling across multiple nodes via the existing gRPC switchboard.
    \item \textbf{Tooling and Ecosystem}: Developing a robust Language Server Protocol (LSP) implementation for advanced IDE support, and a native package manager for distributing Turn modules and schema adapters.
    \item \textbf{Formal Verification}: Extending the formal semantics of Cognitive Type Safety to full mechanical verification of the compiler's schema generation and the VM's validation pipeline.
    \item \textbf{Multi-modal Primitives}: Extending the \texttt{infer} primitive to natively accept image, audio, and video inputs, and exploring broader probabilistic reasoning constructs.
\end{itemize}

\paragraph{Call for Collaboration.} Turn is an ambitious undertaking that sits at the intersection of programming language theory and artificial intelligence. We actively invite researchers, compiler engineers, and open-source contributors to collaborate on these future directions and help shape the foundation of agentic software.

\newpage
\section{Conclusion}
\label{sec:conclusion}

We have presented Turn, a programming language that makes the inherent structure of agentic computation explicit and enforceable. By introducing Cognitive Type Safety, probabilistic control flow, actor-isolated processes, capability-based identity, and compile-time schema absorption as first-class language constructs, Turn eliminates five fundamental failure modes of framework-based agent development.

\begin{tcolorbox}[turnbox,
  colback=red!3,colframe=red!55!black,
  title={Key Insight}]
An agent is a process, not a loop. It has bounded context, persistent memory, a mailbox, and a lifecycle. Encoding these constraints in the language lets the compiler and runtime enforce them mechanically, eliminating the entire class of failures that arise when critical invariants are left to developer discipline.
\end{tcolorbox}

\noindent Turn demonstrates that a small, focused language with the right primitives provides stronger guarantees for agentic software than any library built on a general-purpose host.

Turn is open source under the MIT license. The compiler, VM, standard library, and documentation are available at \url{https://github.com/ekizito96/Turn}. The language specification and interactive examples are at \url{https://turn-lang.dev}.

\clearpage
\twocolumn

% ─────────────────────────────────────────────────────────────────────────────
%  SUPPLEMENTARY MATERIAL
% ─────────────────────────────────────────────────────────────────────────────
\newpage
\onecolumn

\begin{center}
  {\LARGE\bfseries Turn: A Language for Agentic Computation\par}
  \vspace{0.4em}
  {\large Supplementary Material\par}
\end{center}

\vspace{0.6em}

\noindent The supplementary material provides formal depth for the core contributions. Section~\ref{app:opsem} gives the small-step operational semantics for Turn's four novel constructs---\texttt{infer}, \texttt{confidence}, \texttt{grant identity}, and \texttt{spawn\_link}---in a form that complements the typing rules in \S\ref{sec:cognitive-type-safety} and the evaluation in \S\ref{sec:evaluation}. Section~\ref{app:reproducibility} gives exact commands to reproduce all five experimental suites. The complete Turn grammar, worked examples, and tutorials are available at \url{https://turn-lang.dev} and \url{https://github.com/ekizito96/Turn}.

% ─────────────────────────────────────────────────────────────────────────────
\section*{A\quad Extended Operational Semantics}
\label{app:opsem}
\addcontentsline{toc}{section}{A\quad Extended Operational Semantics}

\noindent We give small-step reduction rules for Turn's four novel constructs. The reduction relation $e, \sigma \longrightarrow e', \sigma'$ steps expression $e$ in heap $\sigma$ to $e'$ with updated heap $\sigma'$. We write $\mathcal{S}(T)$ for the JSON Schema generated from struct $T$, $\text{validate}(j, s)$ for JSON Schema validation, and $\mathit{env}(k)$ for the process-environment lookup of key $k$.

\paragraph{Cognitive Type Safety.}
The \textsc{Infer-Step} rule dispatches an inference request carrying the compiled schema; the model response $j$ is validated before binding. If validation fails, \textsc{Infer-Retry} re-prompts (up to $k$ retries); \textsc{Infer-Fail} raises a catchable error when retries are exhausted.

\[
\frac{
  \Gamma \vdash T : \textsf{Struct}
  \quad \mathcal{S}(T) = s
  \quad \text{llm}(e, s) = j
  \quad \text{validate}(j, s) = \textsf{ok}
}{
  \texttt{infer}\;T\;\{e\},\;\sigma
  \;\longrightarrow\;
  \text{parse}(j, T),\;\sigma
}
\;\textsc{(Infer-Ok)}
\]

\[
\frac{
  \text{validate}(j, s) = \textsf{err}(m)
  \quad \text{retries} < k
}{
  \texttt{infer}\;T\;\{e\},\;\sigma
  \;\longrightarrow\;
  \texttt{infer}\;T\;\{e \mathbin{+\!\!+} m\},\;\sigma
}
\;\textsc{(Infer-Retry)}
\]

\[
\frac{
  \text{validate}(j, s) = \textsf{err}(\cdot)
  \quad \text{retries} = k
}{
  \texttt{infer}\;T\;\{e\},\;\sigma
  \;\longrightarrow\;
  \texttt{throw}(\texttt{"InferError"}),\;\sigma
}
\;\textsc{(Infer-Fail)}
\]

\paragraph{Confidence Operator.}

\[
\frac{
  \sigma(v) = \texttt{Uncertain}(w, p)
}{
  \texttt{confidence}\;v,\;\sigma \;\longrightarrow\; p,\;\sigma
}
\;\textsc{(Conf-Uncertain)}
\]
\[
\frac{
  \sigma(v) \neq \texttt{Uncertain}(\cdot,\cdot)
}{
  \texttt{confidence}\;v,\;\sigma \;\longrightarrow\; 1.0,\;\sigma
}
\;\textsc{(Conf-Certain)}
\]

Arithmetic on uncertain values propagates by the product rule; boolean \texttt{and} uses the Zadeh minimum:

\[
\frac{
  \sigma(u) = \texttt{Uncertain}(a, p_1)
  \quad \sigma(v) = \texttt{Uncertain}(b, p_2)
}{
  u \mathbin{+} v,\;\sigma
  \;\longrightarrow\;
  \texttt{Uncertain}(a+b,\; p_1 \cdot p_2),\;\sigma
}
\;\textsc{(Arith-Prop)}
\]

\[
\frac{
  \sigma(u) = \texttt{Uncertain}(\cdot, p_1)
  \quad \sigma(v) = \texttt{Uncertain}(\cdot, p_2)
}{
  u \;\texttt{and}\; v,\;\sigma
  \;\longrightarrow\;
  \texttt{Uncertain}(\cdot,\; \min(p_1, p_2)),\;\sigma
}
\;\textsc{(Bool-Zadeh)}
\]

\paragraph{Capability-Based Identity.}

\[
\frac{
  \mathit{name} \in \Sigma_{\mathit{providers}}
}{
  \texttt{grant identity::}\sigma\texttt{(}\mathit{name}\texttt{)},\;\sigma
  \;\longrightarrow\;
  \texttt{Identity}(\mathit{name}),\;\sigma
}
\;\textsc{(Grant-Ok)}
\]

The \texttt{Identity}$(\mathit{name})$ token carries only the provider name. The raw credential is retrieved by the kernel trap at call time and never materialized as a Turn value:

\[
\frac{
  \begin{array}{c}
  \sigma(h) = \texttt{Identity}(\mathit{name})
  \quad \mathit{env}(\texttt{TURN\_IDENTITY\_}\mathit{name}\texttt{\_TOKEN}) = t \\[2pt]
  \text{http}(\mathit{url},\; t) = r
  \end{array}
}{
  \texttt{\_\_sys\_http\_get}(h,\;\mathit{url}),\;\sigma
  \;\longrightarrow\;
  r,\;\sigma
}
\;\textsc{(Sys-Http)}
\]

Note that $t$ appears only as a parameter to the external $\text{http}$ oracle and is never bound to a Turn variable.

\paragraph{Actor Spawn.}

\[
\frac{
  \mathit{pid} \text{ fresh}
  \quad P_{\mathit{pid}} = (E_{\emptyset},\; C_{\emptyset},\; M_{\emptyset},\; B_{\emptyset},\; \overrightarrow{body})
}{
  \texttt{spawn\_link}\;\texttt{turn()}\;\{body\},\;\sigma
  \;\longrightarrow\;
  \mathit{pid},\;\sigma \uplus \{P_{\mathit{pid}}\}
}
\;\textsc{(Spawn-Link)}
\]

The spawned process receives a fresh tuple $(E_\emptyset, C_\emptyset, M_\emptyset, B_\emptyset, \mathit{pc})$. No component of the parent's state is shared, which is the structural basis for Property~1 (Context Isolation). The bidirectional link means that a \texttt{ProcessExit} signal is delivered to the parent's mailbox if the child terminates abnormally:

\[
\frac{
  P_{\mathit{child}} \;\text{terminates with reason}\; r
  \quad \mathit{parent} \in \text{links}(\mathit{child})
}{
  B_{\mathit{parent}} \;\longleftarrow\; \{\texttt{"type"}: \texttt{"exit"},\; \texttt{"reason"}: r\}
}
\;\textsc{(Exit-Signal)}
\]

% ─────────────────────────────────────────────────────────────────────────────
\section*{B\quad Experiment Reproducibility}
\label{app:reproducibility}
\addcontentsline{toc}{section}{B\quad Experiment Reproducibility}

\noindent All experiments in Section~\ref{sec:evaluation} are fully reproducible from the public repository. The following commands build the compiler and run all five experiment suites on any machine with Rust~1.75+ installed.

\begin{tcolorbox}[colback=gray!5,colframe=gray!50!black,arc=1mm,boxrule=0.5pt,left=4pt,right=4pt,top=4pt,bottom=4pt]
\begin{verbatim}
# Clone and build (release profile matches the paper's measurements)
git clone https://github.com/ekizito96/Turn.git
cd Turn/impl
cargo build --release

# Run all five experiment suites with output
cargo test --release \
  --test experiment_e1_credential_opacity \
  --test experiment_e2_confidence_propagation \
  --test experiment_e3_structured_context \
  --test experiment_e4_memory_scale \
  --test experiment_e5_durable_execution \
  -- --nocapture
\end{verbatim}
\end{tcolorbox}

\noindent Expected output (Apple M4; latencies will vary by hardware):

\begin{tcolorbox}[colback=gray!5,colframe=gray!50!black,arc=1mm,boxrule=0.5pt,left=4pt,right=4pt,top=4pt,bottom=4pt]
\footnotesize
\begin{verbatim}
E1-A PASS: Identity displayed as "<identity stripe>" (opaque, no raw credential)
E1-B PASS: grant identity::oauth("stripe") -> Identity("stripe") (token never in heap)
E1-C PASS: Identity is a certain (non-uncertain) value
E1-D PASS: Two grants for same provider are independent identity handles

E2-1 PASS: confidence(42) = 1.0
E2-2 PASS: confidence(Uncertain(42, 0.73)) = 0.73
E2-3 PASS: confidence(Uncertain(10,0.8) + Uncertain(5,0.5)) = 0.40  (product rule)
E2-4 PASS: confidence(x and y) where x~0.9, y~0.5 = 0.50  (Zadeh min)
E2-5 PASS: confidence=0.45 < 0.70 -> deterministic fallback branch executed
E2-6 PASS: confidence=0.92 >= 0.70 -> high-confidence path executed

E3-1 PASS: P0 system prompt is always first in flat context (primacy position)
E3-2 PASS: most recently appended item is last in flat context (recency position)
E3-3 PASS: P1 overflow evicts oldest item to P2 episodic (demote, not drop)
          P1 size=100, P2 size=1
E3-4 PASS: flat rendering order confirmed as P0(0) -> P2(1) -> P1(101)
E3-5 PASS: each Runtime starts with an empty context (isolation invariant)

E4-1 PASS: K=1000 entries   | write 519 ns/op  | read 574 ns/op
E4-2 PASS: K=10000 entries  | write 1021 ns/op | read 867 ns/op
E4-3 PASS: K=100000 entries | write 594 ns/op  | read 430 ns/op  (O(1) confirmed)
E4-4 PASS: Agent memory is fully isolated -- no cross-process contamination
E4-5 PASS: serialized footprint: 76.8 bytes/entry

E5-1 PASS: mem=10 entries   | state size=875 B (0.9 KB)    | round-trip=211 us
E5-2 PASS: mem=500 entries  | state size=20295 B (19.8 KB)  | round-trip=1340 us
E5-3 PASS: mem=5000 entries | state size=208295 B (203.4 KB)| round-trip=6828 us
E5-4 PASS: full state fidelity after serialize -> deserialize round-trip
\end{verbatim}
\end{tcolorbox}

\noindent To run the complete test suite (24 experiment + 81 unit tests):

\begin{tcolorbox}[colback=gray!5,colframe=gray!50!black,arc=1mm,boxrule=0.5pt,left=4pt,right=4pt,top=4pt,bottom=4pt]
\begin{verbatim}
cargo test --release   # 105/105 pass
\end{verbatim}
\end{tcolorbox}

\end{document}